\documentclass[12pt,draftclsnofoot,onecolumn,doublespace]{IEEEtran}

\IEEEoverridecommandlockouts

\usepackage[cmex10]{amsmath}
\usepackage[dvips]{graphicx}
\usepackage{cite}
\usepackage{amssymb}

\begin{document}
\title{LOS-based Conjugate Beamforming and Power-Scaling Law in Massive-MIMO
Systems
\thanks{Dian-Wu Yue is with the College of Information Science and
Technology, Dalian Maritime University, Dalian, Liaoning 116026,
China (e-mail: dwyue@dlmu.edu.cn). Geoffrey Ye Li is with School
of Electrical and Computer Engineering, Georgia Institute of
Technology, Atlanta, GA 30332-0250, USA
(Email:liye@ece.gatech.edu)} }
\author{Dian-Wu Yue, \emph{Member, IEEE}, and  Geoffrey Ye Li, \emph{Fellow, IEEE}\\
}


\newcommand{\be}{\begin{equation}}
\newcommand{\ee}{\end{equation}}
\newcommand{\bee}{\begin{eqnarray}}
\newcommand{\eee}{\end{eqnarray}}
\newcommand{\nnb}{\nonumber}

\newcommand{\mG}{\mathbf{G}}
\newcommand{\mH}{\mathbf{H}}
\newcommand{\mI}{\mathbf{I}}
\newcommand{\mR}{\mathbf{R}}
\newcommand{\mY}{\mathbf{Y}}
\newcommand{\mZ}{\mathbf{Z}}
\newcommand{\mD}{\mathbf{D}}

\newcommand{\my}{\mathbf{y}}
\newcommand{\mx}{\mathbf{x}}
\newcommand{\mz}{\mathbf{z}}

\newcommand{\mr}{\mathbf{r}}
\newcommand{\mt}{\mathbf{t}}
\newcommand{\mb}{\mathbf{b}}

\newcommand{\mh}{\mathbf{h}}
\newcommand{\mw}{\mathbf{w}}
\newcommand{\mg}{\mathbf{g}}

\newcommand{\mv}{\mathbf{v}}
\newcommand{\ms}{\mathbf{s}}
\newcommand{\mmu}{\mathbf{u}}

\newtheorem{Lemma}{Lemma}
\newtheorem{Theorem}{Theorem}
\newtheorem{Corollary}{Corollary}
\newtheorem{Proposition}{Proposition}
\newtheorem{Example}{Example}
\newtheorem{Definition}{Definition}

\maketitle

\begin{abstract}
This paper is concerned with massive multiple-input
multiple-output (MIMO) systems over Rician flat fading channels.
In order to reduce the overhead to obtain full channel state
information and to avoid the pilot contamination problem, by
treating the scattered component as interference, we investigate a
transmit and receive conjugate beamforming (BF) transmission
scheme only based on the line-of-sight (LOS) component. Under
Rank-1 model, we first consider a single-user system with $N$
transmit and $M$ receive antennas, and focus on the problem of
power-scaling law when the transmit power is scaled down
proportionally to $\frac{1}{MN}$. It can be shown that as $MN$
grows large, the scattered interference vanishes, and the
ergodic achievable rate is higher than that of the corresponding BF scheme
based fast fading and minimum mean-square error (MMSE) channel
estimation. Then we further consider uplink and downlink
single-cell scenarios where the base station (BS) has $M$ antennas
and each of $K$ users has $N$ antennas. When the transmit power
for each user is scaled down proportionally to $\frac{1}{MN}$, it
can be shown for finite users that as $M$ grows without bound,
each user obtains finally the same rate performance as in the
single-user case. Even when $N$ grows without bound, however,
there still remains inter-user LOS interference that can not be
cancelled. Regarding infinite users, there exists such a power
scaling law that when $K$ and $M^\alpha$ go to infinity with a
fixed and finite ratio for a given $\alpha \in (0, 1)$,
 not only inter-user LOS interference but also fast fading
effect can be cancelled,  while fast fading effect can not be
cancelled if $\alpha=1$. Extension to multi-cells and
frequency-selective channels are also discussed shortly. Moreover,
numerical results indicate that spacial antenna correlation does
not have serious influence on the rate performance, and the BS
antennas may be allowed to be placed compactly when $M$ is very
large.
\end{abstract}



\section{Introduction}

Wireless transmission using multiple antennas has attracted much
interest in the past couple of decades due to its capability to
exploit the tremendous capacity inherent in MIMO techniques.
Various aspects of wireless MIMO systems have been studied
intensively, especially the important capacity aspect
\cite{Goldsmith}. Whilst single-user systems have been well
investigated, multi-user systems including classical uplink
(multiple access) and downlink (broadcast) systems nowadays have
become the focus of theoretical analysis and practical design of
MIMO communications \cite{Gesbert}. Theoretically, the
maximum-likelihood multiuser detector and ``dirty paper coding''
can be used to obtain optimal performance for the uplink and
downlink systems, respectively. However, they induce a significant
complexity burden on the system implementation, especially for a
large multiple antenna system. Therefore, linear effective
processing schemes, in particular beam-forming (BF) and
zero-forcing (ZF) detecting or precoding, are of particular
interest as low-complexity alternatives \cite{Peel}
-\cite{Wiesel}.

Recently, there exist a lot of interests in multiuser MIMO with a
very large antenna array at the base station (BS), which means a
array comprising a few hundreds of antennas simultaneously serving
tens of users \cite{Marzetta} -\cite{Lim}. These large scale MIMO
systems can offer much higher data rates, increased link
reliability, and potential power savings since the transmitted RF
energy can be more sharply focused in space while many random
impairments can be averaged out, which is a critical difference
from the traditional MIMO systems. It should be pointed out that
these benefits of large-scale antenna arrays can be reaped by
using the simple BF or ZF processing \cite{Marzetta}, \cite{Yang}.

The analysis and design of massive MIMO systems is at the moment a
fairly new research topic \cite{Rusek}.  In \cite{Gao}, linear
precoding performance is studied for measured very-large MIMO
downlink channels. It is shown that there exist clearly benefits
with an excessive number of BS antennas \cite{Gao}. In \cite{Ngo}
, \cite{Matthaiou} and \cite{Ngo1}, with simple linear BF and ZF
receivers authors give uplink capacity analysis of single-cell,
single-cell distributed, and multi-cell very large MIMO systems,
respectively, derive bounds on the achievable sum rate in both
small and large-scale fading environments, and provide asymptotic
performance results when the number of antennas grows without
bound. In \cite{Hoydis}, authors provide a unified analysis of the
uplink and downlink performance of linear processing in multi-cell
systems when the number of the BS antennas and the number of users
both grow large with a fixed ratio, and derive asymptotically
tight approximations of the achievable rates under a realistic
system model which accounts for channel estimation, pilot
contamination, path loss, and antenna correlation.

In order to achieve the performance predicted by the
mentioned-above analysis results, BS must acquire channel state
information (CSI). In practice, however, the BS does not have
perfect SCI \cite{Ngo3}. Instead, it estimates the channels. The
conventional way of doing this is to use uplink pilots. If channel
coherence time is limited, the number of possible orthogonal pilot
sequences is limited too and hence, pilot sequences have to be
reused in other cells. Therefore, channel estimates obtained in a
given cell will be contaminated by pilots transmitted by users in
other cells \cite{Ngo3}. This causes pilot contamination. The
effect of pilot contamination appears to be a fundamental
challenge of massive-MIMO system design, which warrants future
research on the topic \cite{Marzetta}, \cite{Rusek}, \cite{Yin},
\cite{Ngo2}.

So far there have appeared most of research results on the massive
MIMO based on the ground of Rayleigh fading (see \cite{Marzetta}
-\cite{Ngo2} and references therein among others). However,
despite their practical significance, and in contrast to the
Rayleigh fading case, there are currently very few which apply for
Rician fading, \cite{Zhang1}, \cite{Zhang}, \cite{Liu}. In
particular, authors  in \cite{Zhang1} propose a deterministic
equivalent of ergodic sum rate and an algorithm for evaluating the
capacity-achieving input covariance matrices for uplink massive
MIMO systems. In \cite{Zhang}, authors study the achievable uplink
rates of massive MIMO systems using BF and ZF receivers. Assuming
imperfect CSI, they find such a power scaling law that with a
non-zero Rician factor, the uplink rates converge to the same
constant values when the number of BS antennas $M$ grows large if
the needed transmit power of each user is scaled down
proportionally to $1/M$. However, in pure Rayleigh fading
environment, the corresponding transmit power can only be scaled
down by a factor of $1/\sqrt{M}$ \cite{Ngo}, \cite{Zhang}.

The Rician fading model is also very important and applicable when
the wireless link between the transmitter and the receiver has a
direct path component in addition to the diffused Rayleigh
component. It can be employed in diverse modern applications, like
suburban/indoor WLANs  or 60 GHz communications, to deliver
ultra-broadband data rates \cite{Matthaiou1}. Another emerging
applications are typical point-to-point microwave links and MIMO
vehicular networks, where a moving vehicle communicates with
either another vehicle or with the roadside in support of
demanding applications spanning high-speed networking and video
streaming to mobile commerce and Web surfing \cite{Matthaiou1}.
Moreover, it is also suitable for application in small cell
networks \cite{Hoydis1}. With decreasing cell sizes, the user
terminal are likely to have line-of-sight (LOS) links to one or
several BSs. This means the the normally fasting-fading wireless
channel contains strong deterministic non-fading components
\cite{Hoydis1}, \cite{Zahir}.

For this reason, we are concerned with massive-MIMO systems over
Rician fading channels in this paper. In order to avoid the pilot
contamination problem, we treat the scattered component as
interference, and study a transmit and receive BF transmission
scheme only based on the LOS component. In what follows, we first
consider a single-user MIMO system, and then consider uplink and
downlink scenarios in a single-cell MIMO network.

\section{Single-User Massive-MIMO systems}

\subsection{Single-User System Model}
We consider a single-user MIMO system with $N$ transmit antennas
and $M$ receive antennas.  Then $M \times 1$ received signal
vector is represented as \be  \label{yyy} \my=\sqrt{p_u}\mG\mx+\mz
\ee where  $p_u$ is the average transmitted power of the single
user; $\mG$ is the $M \times N$ channel matrix such that
$[\mG]_{mn}=g_{mn}$, and $g_{mn}$ represents the channel
coefficient between the $m$-th receive antenna and $n$-th transmit
antenna; $\mx$ is the symbol vector transmitted by the user; and
$\mz$ is a vector of zero-mean additive white Gaussian noise
(AWGN) with covariance matrix $\mathbb{E}[\mz\mz^\texttt{H}]=\mI_M$ .

Taking into account the effects of fast fading, geometric
attenuation, and shadow fading, the entry $g_{mn}$ of $\mG$ should
be modelled as \be  g_{mn}=h_{mn}\sqrt{\beta} \ee where $h_{mn}$
represents the fast fading coefficient from the $n$-th transmit
antenna to the $m$-th receive antenna, and $\beta$ models the
geometric attenuation and shadow fading, which is independent over
$m$ and $n$. Therefore, we have \be\mG=\sqrt{\beta}\mH \ee where
$\mH=[h_{mn}]$.

We assume that the fast fading is Rician frequency-flat. Then the
matrix $\mH$ can be decomposed into a sum of a specular matrix and
a scattered matrix, i.e., \be \label{hhh}
\mH=\sqrt{\bar{\vartheta}}\overline{\mH}+\sqrt{\tilde{\vartheta}}\widetilde{\mH}
\ee where $\bar{\vartheta}=\frac{\vartheta}{1+\vartheta}$,
$\tilde{\vartheta}=\frac{1}{1+\vartheta}$, and $\vartheta \geq 0$
is just the Rician $K$-factor. Note that $\vartheta=0$ corresponds
to a Rayleigh fading while $\vartheta \rightarrow \infty$
corresponds to non-fading channels. In this paper, we always
assume that $\vartheta>0$. The specular matrix $\overline{\mH}$ in
(\ref{hhh}) is given by \cite{Farrokhi} -\cite{Taricco} \be
\overline{\mH}=\mr\mt^\texttt{T} \ee where $\mr$ and $\mt$ are the
specular array responses at receiver and transmitter,
respectively, and can be written as \be \mr=[1, e^{j2\pi d_r
\sin(\theta)}, \ldots, e^{j2\pi (M-1)d_r \sin(\theta)}]^\texttt{T}
\ee and \be \mt=[1, e^{j2\pi d_t \sin(\phi)}, \ldots, e^{j2\pi
(N-1)d_t \sin(\phi)}]^\texttt{T} \ee where $\theta$ or $\phi$ is
the angle of arrival or departure of the specular component, and
$d_r$ or $d_t$ is the antenna spacing in wavelengths at receiver
or transmitter. For the scattered matrix $\widetilde{\mH}$ in
(\ref{hhh}), its entries are independent and identically
distributed (i.i.d) circular complex Gaussian random variables
with zero mean unit variance, i.e., $\widetilde{h}_{mn} \sim
\texttt{CN}(0,1).$

Through the paper, we assume that neither the transmitter nor the
receiver knows the scattered component, but both of them knows the
specular component.

\subsection{Transmit/Receive BF Scheme based on the Specular Component}
A MIMO system can be configured differently. One configuration is
transmit/receive BF which has been widely used due to its
simplicity and good performance. It is well known that traditional
transmit/receive BF schemes are usually based on the scattered
component. In this paper, we investigate a transmit/receive BF
scheme only based on the specular component due to the assumption
which neither the transmitter nor the receiver knows the scattered
component.

Transmit/receive BF system transmits one symbol at a time. We
denote by $s$ the desired information symbol such that
$\mathbb{E}s^Hs=1$. Then $s$ is first weighted by the transmit
beamformer $\mb$ with $\mathbb{E}\mb^\texttt{H}\mb=1$ before being
feeded to the $N$ transmit antennas. Now let \be
\mg=\mG\mb=\sqrt{\bar{\vartheta}\beta}\bar{\mh}
+\sqrt{\tilde{\vartheta}\beta}\tilde{\mh}=\bar{\mg}+\tilde{\mg}
\ee where $\bar{\mh}=\bar{\mH}\mb$ and
$\tilde{\mh}=\tilde{\mH}\mb$. Then the received vector in
(\ref{yyy}) can be rewritten as \bee \label{yyy1}
\my&=&\sqrt{p_u}\mg s+\mz \nnb
\\ &=&\sqrt{p_u\bar{\vartheta}\beta}\bar{\mh}
s+\sqrt{p_u\tilde{\vartheta}\beta}\tilde{\mh} s +\mz\nnb
\\ &=& \sqrt{p_u\bar{\vartheta}\beta}\bar{\mh}
s+\bar{\mz}\eee where
$\bar{\mz}=\sqrt{p_u\tilde{\vartheta}\beta}\tilde{\mh} s +\mz$.
(\ref{yyy1}) implies that the considered MIMO system can be viewed
as a MIMO system operating in a pure LOS environment with additive
noise $\bar{\mz}$.

After receiving $\my$, the receiver employs a weighting vector
$\mw$ to combine $\my$ to a single decision variable. The transmit
and receive weighting vectors, $\mb$ and $\mw$, should be chosen
to maximize the output signal-to-interference-noise ratio (SINR),
as defined by \cite{Yue} \be \label{gg}
\gamma=p_u\beta\bar{\vartheta}\cdot\frac{\mw^\texttt{H}
(\bar{\mH}\mb)(\bar{\mH}\mb)^\texttt{H}\mw}{\mw^\texttt{H}\Omega\mw}\ee
where \bee
\Omega&=&\mathbb{E}\bar{\mz}\bar{\mz}^\texttt{H}\nnb\\
&=&\mathbb{E}(\sqrt{p_u\beta\tilde{\vartheta}}\tilde{\mh}s+\mz)
(\sqrt{p_u\beta\tilde{\vartheta}}\tilde{\mh}s+\mz)^\texttt{H}\nnb\\
&=&p_u\beta\tilde{\vartheta}\tilde{\mh}\tilde{\mh}^\texttt{H}+\mI_M.\eee
It should be noticed that neither the transmitter nor the
receiver can know $\Omega$, but both of them can know its statistical average $\bar{\Omega}$.
Since the $M \times 1$
random vector $\tilde{\mh}$ follows a complex Gaussian
distribution of mean zero and covariance matrix
$\mathbb{E}[\tilde{\mh}\tilde{\mh}^\texttt{H}]=\mI_M$, thus \be\bar{\Omega}=(p_u\beta\tilde{\vartheta}+1)\mI_M.\ee
In (\ref{gg}), we replace $\Omega$ by $\bar{\Omega}$ and denote the new expression by $\bar{\gamma}$. Then
\be \label{gg1} \bar{\gamma}=\frac{p_u\beta\bar{\vartheta}}{p_u\beta\tilde{\vartheta}+1}\cdot \frac{\mw^\texttt{H}
(\bar{\mH}\mb)(\bar{\mH}\mb)^\texttt{H}\mw}{\mw^\texttt{H}\mw}. \ee
$\bar{\gamma}$ is an important parameter in the whole paper. We refer to it as the output statistical SINR, and will carry on our analysis with its help.

Based on $\bar{\gamma}$ rather than $\gamma$,  the optimum weighting vectors
$\mb$ and $\mw$ can be chose as \cite{Kang} \be
\mb=\frac{\mt}{\sqrt{N}}, \;\; \mw=\frac{\mr}{\sqrt{M}}. \ee
So we have the following result.

\begin{Proposition} \label{P1} The statistical SINR is given by\be
\bar{\gamma}=\frac{NMp_u\beta\bar{\vartheta}}{1+p_u\beta\tilde{\vartheta}}
\ee
\end{Proposition}

Through analysis in asymptotic cases,  Proposition \ref{P1} has
the following three corollaries.

\begin{Corollary} \label{C1}
If $MN$ and $\vartheta$ are fixed, and $p_u \to \infty$, then \be
\bar{\gamma} \to MN \vartheta \ee
\end{Corollary}

\begin{Corollary} \label{C2}
If $MN$, $\beta$ and $p_u$ are fixed, and $\vartheta \to \infty $,
then \be \bar{\gamma} \to MN \beta p_u \ee
\end{Corollary}

\begin{Corollary} \label{C3}
If $\vartheta$ and $\beta$ are fixed, and let $E_u=MNp_u$ be fixed
and $MN \to \infty$, then \be \bar{\gamma} \to E_u
\beta\bar{\vartheta}.\ee
\end{Corollary}

We denote the ergodic achievable rate by $ R=\mathbb{E}\log_2(1+\gamma)$. Then we have the following lemma about $R$.
\begin{Theorem}
\be  \log_2 (1+\bar{\gamma}) \leq R \leq  \log_2(1+\bar{\gamma}^\texttt{L})\ee
where $\bar{\gamma}^\texttt{L}=E_u\beta\bar{\vartheta}$. Furthermore, if $E_u=MNp_u$ is fixed when $MN \to \infty$, then
\be \label{tt} \lim_{MN \to \infty} R=\lim_{MN \to \infty} \bar{R}=\bar{R}^\texttt{L}\ee
where $\bar{R}=\log_2 (1+\bar{\gamma})$ and $\bar{R}^\texttt{L}=\log_2(1+\bar{\gamma}^\texttt{L}).$
\end{Theorem}
{\em Proof:} By using Jensen's inequality, we obtain the following lower bound on the ergodic achievable rate
\bee R &\geq& \log_2(1+\frac{1}{\mathbb{E}(1/\gamma)})\nnb \\
 &=& \log_2(1+\frac{p_u\beta\bar{\vartheta}MN}{p_u\tilde{\vartheta}\mathbb{E}|\mw^H\tilde{\mh}|^2+\mathbb{E}|\mw^H\mz|^2}). \eee
Since $\mathbb{E}|\mw^H\mz|^2=1$ and $\mathbb{E}|\mw^H\tilde{\mh}|^2=1$, we have further
\be  R \geq  \log_2 (1+\frac{p_u\beta\bar{\vartheta}MN}{1+p_u\beta\tilde{\vartheta}})=\log_2 (1+\bar{\gamma}). \ee
On the other hand, it follows easily that
\bee R&=& \mathbb{E}\log_2 ( \frac{p_u\beta\bar{\vartheta}MN}{p_u\beta
\tilde{\vartheta}|\mw^H\tilde{\mh}|^2+1})\nnb\\
     &\leq& \mathbb{E} \log_2(1+p_u\beta\bar{\vartheta}MN )\nnb\\
     &=&\log_2(1+\bar{\gamma}^\texttt{L}). \eee
If  $E_u=MNp_u$ is fixed when $MN \to \infty$, by using the upper and lower bounds we easily obtain the desired result (\ref{tt}).
\;\; $\blacksquare$

Theorem 1 reveals such a power scaling law that without
degradation in the rate performance, the transmit power can be cut
down by a factor of $\frac{1}{MN}$ when $MN$ grows large.

\subsection{Receive BF Scheme based on FF and MMSE Channel Estimation}

In fact,  we have just presented such a simple transmit/receive BF scheme
that the transmitter and receiver only need to know
partial information about the LOS component $\mt$ and $\mr$,
respectively.  From another point of view, (\ref{yyy1}) also
reveals that the considered MIMO system through the transmit BF is
equivalent to a SIMO system. For this reason, using (\ref{yyy1})
as a starting point, we can also develop a receive BF (or say
maximum-ratio combining (MRC)) scheme based on fast fading (FF)
and minimum mean-squared error (MMSE) channel estimation by
following the research ideas in \cite{Ngo} and \cite{Zhang}. The
presupposition is that the receiver can have imperfect information
about the scattered component (or say FF).

For the mentioned-above FF-based receive BF scheme, the equivalent
channel vector $\mg$ needs to be estimated at receiver by using
pilots from the transmitter. Its specular component $\bar{\mg}$
(including $\beta$, $\vartheta$, and $\bar{\mh}$) is assumed to be
constant over many coherence time intervals and known a priori.
Therefore, only its scattered component $\tilde{\mg}$ needs to be
estimated.

Now let $T$ denote the length of the coherence interval and $\tau<
T$ denote the number of symbols used for the pilots.   The used
pilot sequence can be represented by a $\tau \times 1$ vector
$\sqrt{p_{\texttt{p}}} \mathbf{\Phi } $ satisfying
$\mathbf{\Phi}^\texttt{H}\mathbf{\Phi}=1$, with $
p_{\texttt{p}}=\tau p_u$. Then the $M \times \tau$ received pilot
matrix at the receiver is given by \be
\mY_{\texttt{p}}=\sqrt{p_{\texttt{p}}}\mg\mathbf{\Phi}^\texttt{T}+\mZ_{\texttt{p}}
\ee where $\mZ_{\texttt{p}}=(z_{ij}^{\texttt{p}})$ is an $M \times \tau$ matrix
whose elements $\{ z_{ij}^{\texttt{p}}\}$ are i.i.d., and
$z_{ij}^{\texttt{p}}\sim\texttt{CN}(0,1)$. If let
$\tilde{\mY}_{\texttt{p}}=\mY_{\texttt{p}}-\sqrt{p_{\texttt{p}}}\bar{\mg}\mathbf{\Phi}^\texttt{T}$,
then MMSE estimate of $\tilde{\mg}$ is given by \cite{Ngo},
\cite{Hassibi} \be
\hat{\tilde{\mg}}=\frac{\sqrt{p_{\texttt{p}}}\beta\tilde{\vartheta}}{1+p_{\texttt{p}}\beta\tilde{\vartheta}}\tilde{\mY}_{\texttt{p}}\mathbf{\Phi}^*
.\ee  Note that $(\cdot)^*$ denotes complex conjugate.  We denote
by $\mathbf{\Xi}=\hat{\tilde{\mg}}-\tilde{\mg}$. By a derivation
similar to that in \cite{Ngo}, it can conclude that the random
vector $\mathbf{\Xi}=(\xi_{i})$ is independent of $\hat{{\mg}}$,
and $\xi_i \sim \texttt{CN}(0,\delta^2_\xi)$, where
$\delta^2_\xi=\frac{\beta\tilde{\vartheta}}{1+p_{\texttt{p}}\beta\tilde{\vartheta}}$.
Moreover, let $\hat{\tilde{\mg}}=(\hat{\tilde{g}}_i)$, and it
follows easily that $\hat{\tilde{g}}_i \sim
\texttt{CN}(0,\delta^2_g)$, where
$\delta^2_g=\frac{\beta^2\tilde{\vartheta}^2p_{\texttt{p}}}{1+p_{\texttt{p}}\beta\tilde{\vartheta}}.$

After MMSE channel estimation, the weighting vector can be chose
as $\mw=\hat{\mg}$. Then the combined signal at the receiver is
given by
\bee  \hat{v}&=&\hat{\mg}^\texttt{H}(\sqrt{p_u}\mg s+\mz) \nnb\\
&=& \sqrt{p_u} \hat{\mg}^\texttt{H}  \hat{\mg} s-\sqrt{p_u}
\hat{\mg}^\texttt{H}  \mathbf{\Xi} s+\hat{\mg}^\texttt{H}\mz. \eee
At this time, the received SINR becomes   \be
\hat{\gamma}=\frac{p_u\|\hat{\mg}\|^2}{p_u\delta^2_\xi+1}. \ee
Accordingly, the achievable rate for the  BF scheme based on FF
and MMSE channel estimation can be written as \be
\hat{R}=\frac{T-\tau}{T}\mathbb{E}\log_2(1+\frac{p_u\|\hat{\mg}\|^2}{p_u\delta^2_\xi+1}).
\ee

\begin{Theorem} \label{P2} If $\vartheta$ and $\beta$ are fixed, and $E_u=MNp_u$
is also fixed when $MN \to \infty$, then \be \lim_{MN \to
\infty}\hat{R} =\frac{T-\tau}{T} \lim_{MN \to \infty}
R=\frac{T-\tau}{T} \lim_{MN \to \infty}
\bar{R}=\frac{T-\tau}{T} \bar{R}^\texttt{L}. \ee
\end{Theorem}

{\em Proof:} With the help of the well-known law of large number,
this result can be derived directly by using the following
expressions
 \be
 \delta^2_\xi=\frac{\beta\tilde{\vartheta}}{1+p_{\texttt{p}}\beta\tilde{\vartheta}},
 \ee \be  \mathbb{E}\|\bar{\mg}\|^2=MN\beta\bar{\vartheta},  \ee \be
\mathbb{E}\|\hat{\tilde\mg}\|^2=\frac{M\beta^2\tilde{\vartheta}^2p_{\texttt{p}}}{1+p_{\texttt{p}}\beta\tilde{\vartheta}}, \ee and \be
\mathbb{E}\|\hat{\mg}\|^2=
\mathbb{E}\|\bar{\mg}\|^2+\mathbb{E}\|\hat{\tilde\mg}\|^2
=MN\beta\bar{\vartheta}+\frac{M\beta^2\tilde{\vartheta}^2p_{\texttt{p}}}{1+p_{\texttt{p}}\beta\tilde{\vartheta}}.\ee
Note that $p_{\texttt{p}} \to 0$ when $MN \to \infty$. \;\; $\blacksquare$

This theorem implies that the ergodic achievable rate of the LOS-based
scheme can be higher than that of the FF-based scheme when $M$ or
$N$ is very large.

\section{Uplink Single-Cell Massive-MIMO systems}

In Section II, we have considered a single-user massive-MIMO
system. This simplifies the rate analysis, and it gives us
important insights into how power can be scaled with the numbers
of transmit and receive antennas. A natural problem is to what
extent this power-scaling law still holds for multi-user
Massive-MIMO systems. In this section, we will consider the
problem and focus on the uplink single-cell scenario.

\subsection{Uplink System Model}

Let us consider an uplink single-cell MIMO system where the BS is
equipped with $M$ antennas and serves $K$ users with each
connected to $N$ antennas.  Then $M \times 1$ received signal
vector at the BS is represented as \be
\my=\sqrt{p_u}\mathbb{G}\mathbb{X}+\mz \ee where $p_u$ is the
average transmitted power of each user; $\mathbb{G}$ is the $M
\times NK$ channel matrix between the BS and the $K$ users;
$\mathbb{X}$ is the symbol vector simultaneously transmitted by
the $K$ users; and $\mz$ still denotes  a zero-mean AWGN vector
with $\mathbb{E}[\mz\mz^\texttt{H}]=\mI_M$.

The channel matrix $\mathbb{G}$ consists of $K$ sub-matrices as
follows: \be     \mathbb{G}=[\mG_1,\mG_2, \ldots, \mG_K] \ee where
the $k$-th sub-matrix $\mG_k$ is described as in Section II. A,
i.e., \be \label{ggg} \mG_k=\sqrt{\beta_k}\mH_k \ee where \be
\label{hhh1}
\mH_k=\sqrt{\bar{\vartheta}_k}\overline{\mH}_k+\sqrt{\tilde{\vartheta}_k}\widetilde{\mH}_k
\ee with  $\bar{\vartheta}_k=\frac{\vartheta_k}{1+\vartheta_k}$,
$\tilde{\vartheta}_k=\frac{1}{1+\vartheta_k}$, and $\vartheta_k
\geq 0$ being the Rician $K$-factor of the $k$-th user. The
entries of $\widetilde{\mH}_k$ are still modelled as i.i.d.
$\texttt{CN}(0,1)$ random variables. And $\overline{\mH}_k$ is
expressed as \label{ee} \be
\overline{\mH}_k=\mr_k\mt_k^\texttt{T}. \ee In (\ref{ee}),
$\mr_k$ and $\mt_k$ are written as \be \label{rrr} \mr_k=[1,
e^{j2\pi d \sin(\theta_k)}, \ldots, e^{j2\pi (M-1)d
\sin(\theta_k)}]^\texttt{T}  \ee and \be \label{ttt1} \mt_k=[1,
e^{j2\pi d_k \sin(\phi_k)}, \ldots, e^{j2\pi (N-1)d_k
\sin(\phi_k)}]^\texttt{T}   \ee where $d$ and $d_k$ are the
antenna spacings in wavelengths at the linear antenna arrays of
the BS and the $k$ user terminals, respectively, while $\theta_k$
and $\phi_k$ are the angles of arrival and departure of the
channel specular component for the $k$-th user, respectively.  We
assume that $\theta_i \neq \theta_k$ when $i\neq k$.

\subsection{Transmit/Receive BF Scheme based on Specular
Component}

The transmit/receive BF scheme based on specular component
presented in Section II can be naturally extended to the uplink
scenario.

For the $k$-th user, as in the single-user case, its transmit
beamformer can be chosen as \be \mb_k=\frac{\mt_k}{\sqrt{N}}, \;\;
k=1,2,\ldots, K. \ee So the transmitted vector $\mathbb{X}$ can be
written as
\bee  \mathbb{X}&=&[\mx_1,\mx_2, \ldots, \mx_K]^\texttt{T} \nnb \\
&=&[\mb_1s_1,\mb_2s_2, \ldots, \mb_Ks_K]^\texttt{T}\eee where
$s_k$ is the symbol transmitted by the $k$-th user such that
$\mathbb{E}s^H_ks_k=1$. At the BS the received vector can be
rewritten as \be \label{yyy2} \my=\sqrt{p_u}\mG\ms+\mz
\ee where $ \ms=[s_1,s_2, \ldots, s_k] $, and
 \be  \mG=[\mg_1,\mg_2, \ldots, \mg_K]                 \ee
with $\mg_k=\mG_k\mb_k$.

(\ref{yyy2}) indicates that the uplink system through the
processing of transmit BF can be reduced to a traditional uplink
system with $K$ single-antenna users, as discussed in \cite{Ngo}
and \cite{Zhang}. Due to this reason, using (\ref{yyy2}) as
a starting point, we can have three well-known linear detectors MRC,
ZF and MMSE available, even under the constraint that the BS only
knows the specular component of channel matrix $\bar{\mG}$. Now
let $\mathbf{\Lambda}$ denote an $M \times K$ linear detecting
matrix which only depends on $\bar{\mG}$. Then these three linear
detectors MRC, ZF and MMSE can be expressed as \cite{Ngo}, \be
\label{lll} \mathbf{\Lambda}=\left\{\begin{array}{ll}
     \bar{\mG}  &  \mbox{for} \; \mbox{MRC};\\
     \bar{\mG}(\bar{\mG}^\texttt{H}\bar{\mG})^{-1} & \mbox{for} \; \mbox{ZF};\\
\bar{\mG}(\bar{\mG}^\texttt{H}\bar{\mG}+(\sum_{k=1}^K\beta_k\tilde{\vartheta}_k+p_u^{-1})\mI_K)^{-1}&
\mbox{for} \; \mbox{MMSE}.
      \end{array}
     \right.\ee As a matter of fact, these three detectors are equivalent to each
other when $K=1$.

After using the linear detector $\mathbf{\Lambda}$,  the received
vector $\my$ becomes \be
\mv=\mathbf{\Lambda}^\texttt{H}(\sqrt{p_u}\mG \ms+\mz). \ee Let
$\mathbf{\lambda}_k$ is the $k$-th element of $\mathbf{\Lambda}$.
Then, the $k$-th element of $\mv$ is written in detail as \be
v_k=\sqrt{p_u}\mathbf{\lambda}_k^\texttt{H}\bar{\mg}_ks_k+\sqrt{p_u}\mathbf{\lambda}_k^\texttt{H}\tilde{\mg}_ks_k+\sqrt{p_u}\sum_{i=1,i\neq
k}^K\mathbf{\lambda}_k^\texttt{H}(\bar{\mg}_i+\tilde{\mg}_i)s_i
+\mathbf{\lambda}_k^\texttt{H}\mz. \ee Thus the output statistical SINR is
given by \be
\bar{\gamma}_k(K)=\frac{p_u|\mathbf{\lambda}_k^\texttt{H}\bar{\mg}_k|^2}{p_u\mathbb{E}|\mathbf{\lambda}_k^\texttt{H}\tilde{\mg}_k|^2+p_u\sum_{i=1,i\neq
k}^K\mathbb{E}|\mathbf{\lambda}_k^\texttt{H}(\bar{\mg}_i+\tilde{\mg}_i)|^2
+\mathbb{E}|\mathbf{\lambda}_k^\texttt{H}\mz|^2}. \ee

\begin{Proposition} \label{P3}
For the MRC detector, we have \be \label{ggg1}
\bar{\gamma}_k(K)=\frac{p_uMN\beta_k\bar{\vartheta}_k}{1+\frac{p_uN}{M}\sum_{i=1,i\neq
k}^K\beta_i\bar{\vartheta}_i|\varrho_{ki}|^2+p_u\sum_{i=1}^K\tilde{\vartheta}_i\beta_i}
\ee where \be
\varrho_{ki}=\frac{1-e^{jM\varphi_{ki}}}{1-e^{j\varphi_{ki}}},
\;\;\varphi_{ki}=2\pi d (\sin(\theta_i)-\sin(\theta_k)). \ee
\end{Proposition}

{\em Proof:} For the MRC detector,
$\mathbf{\lambda}_k=\bar{\mg}_k$. This result is easily derived by
using the following expressions \be
|\mathbf{\lambda}_k^\texttt{H}\bar{\mg}_k|^2=(MN\beta_k\bar{\vartheta}_k)^2,
\ee \be
\mathbb{E}|\mathbf{\lambda}_k^\texttt{H}\mz|^2=MN\beta_k\bar{\vartheta}_k,\ee
\be
\mathbb{E}|\mathbf{\lambda}_k^\texttt{H}\tilde{\mg}_i|^2=MN\beta_k\bar{\vartheta}_k\tilde{\vartheta}_i\beta_i,
\ee and \bee
|\mathbf{\lambda}_k^\texttt{H}\bar{\mg}_i|^2&=&|\sqrt{N\beta_k\bar{\vartheta}_k}\sqrt{N\beta_i\bar{\vartheta}_i}\mr_k^H\mr_i|^2
\nnb\\ &=& N^2 \beta_k\bar{\vartheta}_k\beta_i\bar{\vartheta}_i
|\frac{1-e^{jM\varphi_{ki}}}{1-e^{j\varphi_{ki}}}|^2 \nnb \\
&=& N^2
\beta_k\bar{\vartheta}_k\beta_i\bar{\vartheta}_i|\varrho_{ki}|^2.\eee
\;\;$\blacksquare$

Note that when $MN$ grows large, $|\varrho_{ki}|^2 \leq
\frac{4}{|1-e^{j\varphi_{ki}}|^2}$ still keeps finite. Thus, \be
\lim_{M \to \infty}\frac{1}{M}|\varrho_{ki}|^2=0.   \ee So in a
LOS environment, we also have the following favorable propagation
condition as if in a rich scattering environment
\cite{Marzetta},\cite{Ngo} \be \label{mmm}\lim_{M \to
\infty}\frac{1}{M}\bar{\mG}^\texttt{H}\bar{\mG}=\mD \ee where \be
\mD=\mbox{diag}(\beta_1\bar{\vartheta}_1,
\beta_2\bar{\vartheta}_2, \ldots, \beta_K\bar{\vartheta}_K).\ee

Furthermore, it follows from (\ref{mmm}) and (\ref{lll}) that when
$M$ grows large the ZF and MMSE detectors can tend to that of the
MRC. In reality, however, the implementation of the ZF and MMSE
detectors involve a relatively complicated computation of finding
the inverse of a large dimensional matrix, comparing with the MRC
detector. Therefore, in what follows we focus on the simplest MRC
detection scheme for the asymptotic statistical SINR analysis. We obtain the
following results.

\begin{Corollary} \label{C4} Let $E_u=MNp_u$
be fixed when $M \to \infty$. Then \be \label{mmm2} \lim_{M \to
\infty}\bar{\gamma}_k(K)=\lim_{M \to
\infty}\bar{\gamma}_k(1)=E_u\beta_k\bar{\vartheta}_k. \ee
\end{Corollary}

This corollary shows that the uplink system with multi-users has
the same SINR limit as the one with single-user when the number of
the BS antennas grows without bound. In other words, the very
large antenna array deployed at the BS can eliminate intra-cell
LOS interference and FF impact.

\begin{Corollary} \label{C5} When $N \to \infty$, let $E_u=MNp_u$ be fixed, $M$ be finite, and
$K>1$ be fixed. Then \be \lim_{N \to
\infty}\bar{\gamma}_k(K)=\frac{E_u\beta_k\bar{\vartheta}_k}{1+E_uM^{-2}c(K)}<\lim_{N
\to \infty}\bar{\gamma}_k(1)=E_u\beta_k\bar{\vartheta}_k. \ee
where $c(K)=\sum_{i=1,i\neq
k}^K\beta_i\bar{\vartheta}_i|\varrho_{ki}|^2$.
\end{Corollary}

This corollary implies that even though each user has a very large
antenna array , the intra-cell LOS interference can not be
mitigated while the FF impact can be eliminated.

Regarding (\ref{ggg1}), we define \be \mathbb{E}\beta_i=\beta,\ee
\be \mathbb{E}\bar{\vartheta}_i=\bar{\vartheta}, \ee \be
\mathbb{E}\tilde{\vartheta}_i=\tilde{\vartheta},\ee and \be
\mathbb{E}|\varrho_{ki}|^2=|\varrho_{k}|^2.\ee When $K$ is very
large, the SINR in (\ref{ggg1}) can be approximated as \be
\bar{\gamma}_k(K)\approx\frac{p_uMN\beta_k\bar{\vartheta}_k}{1+\frac{p_uN}{M}(K-1)\beta\bar{\vartheta}|\varrho_{k}|^2+p_uK\tilde{\vartheta}\beta
} \ee

\begin{Corollary} \label{C6} Let $E_u=MNp_u$ be fixed. When $M \to \infty$ and $K \to \infty$, let $ \frac{K}{M^\alpha} \to \iota$, and $\iota>0$.
If $0<\alpha<1$,   then \be \lim_{M \to
\infty}\bar{\gamma}_k(K)=\lim_{M \to
\infty}\bar{\gamma}_k(1)=E_u\beta_k\bar{\vartheta}_k. \ee And if
$\alpha=1$ and $N$ be finite, then \be \lim_{M \to
\infty}\bar{\gamma}_k(K)=\frac{E_u\beta_k\bar{\vartheta}_k}{1+E_u\iota
\tilde{\vartheta}\beta/N}<\lim_{M \to
\infty}\bar{\gamma}_k(1)=E_u\beta_k\bar{\vartheta}_k. \ee
\end{Corollary}

Corollary \ref{C6} indicates that the massive MIMO system by
employing the transmit/receive BF scheme can also serve an
infinite number of users. In particular, the very large antenna
array at the BS can eliminate not only intra-cell LOS interference
but also the FF impact as long as $K$ and $M^\alpha (0<\alpha<1)$
grow with fixed ratio. If $\alpha=1$, however, the system is still
influenced by FF.

For the uplink system with $K>1$ users, as in Section II, we can
also present a receive BF scheme based on FF and MMSE channel
estimation. When using the BF scheme, the ergodic achievable rate for the
$k$-th user can be defined as \be
\hat{R}_k(K)=\frac{T-\tau}{T}\mathbb{E}\log_2(1+\hat{\gamma}_k(K))
\ee where $\hat{\gamma}_k(K)$ denotes the receive SINR at the BS
for the $k$-th user, $T$ represents the coherence time of the
channel in the terms of number of symbols and $\tau$ is the number
of symbols used as pilots for the MMSE channel estimation
\cite{Zhang}. On the other hand, if the system employs the
LOS-based BF scheme, the corresponding achievable rate $R_k(K)$ has the following lower bound:
 \bee \label{rrr1}
\bar{R}_k(K)&=&\log_2(1+\frac{p_u\|\bar{\mg}_k\|^4}{p_u
\mathbb{E}|\bar{\mg}_k^{\texttt{H}}\tilde{\mg}_k|^2+p_u\sum_{i=1,i\neq
k}^K\mathbb{E}\|\bar{\mg}_k(\bar{\mg}_i+\tilde{\mg}_i)\|^2+\|\bar{\mg}_k\|^2})\nnb
\\ &=&\log_2(1+\bar{\gamma}_k(K)). \eee
Moreover, if $E_u=MNp_u$ is fixed when $M \to \infty$, the upper bound of $R_k(K)$ can be
represented as
\be \label{mmm1} \bar{R}^{\texttt{L}}_k=\lim_{M \to \infty}
\bar{R}_k(K)=\log_2(1+E_u \beta_k\bar{\vartheta}_k). \ee

\begin{Theorem} \label{P4} Let $E_u=MNp_u$
be fixed when $M \to \infty$.  Then \be  \lim_{M \to
\infty}\hat{R}_k (K)=\frac{T-\tau}{T} \lim_{M \to \infty}
\bar{R}_k (K)=\frac{T-\tau}{T}\bar{R}^{\texttt{L}}_k.\ee
\end{Theorem}

{\em Proof:} By Theorem \ref{P2} and Expressions (\ref{mmm2}),
(\ref{rrr1}), and (\ref{mmm1}),  it follows that \be \lim_{M \to
\infty}\hat{R}_k(K) \leq \lim_{M \to \infty}\hat{R}_k(1)=\lim_{M
\to \infty} \frac{T-\tau}{T} \bar{R}_k (1)=\lim_{M \to \infty}
\frac{T-\tau}{T} \bar{R}_k
(K)=\frac{T-\tau}{T}\bar{R}^{\texttt{L}}_k.\ee  Then we only need
to prove $\lim_{M \to \infty}\hat{R}_k(K) \geq \lim_{M \to \infty}
\frac{T-\tau}{T} \bar{R}_k (K)$.

Regarding the receive BF scheme based on FF and MMSE channel
estimation, there exists such an extreme case where there is no
pilot sequence to be used to estimate the scattered channel, i.e.,
$\tau=0$. In this case,  the BS will treat the LOS component as
the channel estimate. This means that the LOS-based BF scheme can
be viewed as a special case of the FF-based BF scheme. For this
reason, we have, \be \bar{R}_k (K)=\hat{R}_k(K)|_{\tau=0}. \ee
Moreover, it can follow that \be \hat{R}_k(K)|_{\tau=0} \leq
\frac{T}{T-\tau}\hat{R}_k(K)|_{\tau>0}. \ee Thus \be
\frac{T-\tau}{T} \bar{R}_k (K) \leq \hat{R}_k(K). \ee So we can
have the desired result \be \lim_{M \to \infty}\hat{R}_k
(K)=\lim_{M \to \infty} \frac{T-\tau}{T} \bar{R}_k (K).\ee
Namely, Theorem \ref{P4} holds. \;\; $\blacksquare$

This theorem shows that for the uplink system with multi-users,
the individual ergodic achievable rate with the LOS-based BF scheme can be higher
than that with the FF-based scheme when the number of BS antennas
is very large.

\section{ Downlink Scenario and Other Extension Consideration}

\subsection{Downlink Single-Cell Massive-MIMO}

We next turn our attention to the downlink scenario in which the
BS has $M$ antennas and each of the $K$ users has $N$ antennas. We
use $\mG^\texttt{T}_k$ to denote the downlink channel matrix from
the BS to the $k$-th user such that $\mG_k$ can be described in
(\ref{ggg}). Then the received signal vector for the $k$-th user
can be written as \be \my_k=\sqrt{p_b}\mG_k^\texttt{T}\mathbb{X}+\mz_k \ee
where $p_b$ is the average transmitted power of the BS,
$\mathbb{X}$ is the symbol vector transmitted by the BS, and
$\mz_k$  denotes a zero-mean AWGN vector with
$\mathbb{E}[\mz_k\mz^\texttt{H}_k]=\mI_N$.

For the downlink scenario, we also employ the conjugate BF
precoding rather than ZF or MMSE precoding since the corresponding signal
processing can be done distributedly at each antenna separately.
When making use of the conjugate beamforming precoder, the
transmitted vector can be given by \be  \mathbb{X}=\sum_{k=1}^K
\frac{\mr_ks_k}{\sqrt{KM}} \ee where $\mr_k$ is defined in
(\ref{rrr}), but different from the uplink scenario $\theta_k$ now
denotes the angle of departure of the channel specular component
while $\phi_k$ in (\ref{ttt1}) accordingly becomes the angle of
arrival. The $k$-th user obtains by using the combiner
$\frac{\mt^\texttt{H}_k}{\sqrt{N}}$  \be
v_k=\frac{\mt^\texttt{H}_k}{\sqrt{N}}(\sqrt{p_b}\mG_k^\texttt{T}\mathbb{X}+\mz_k)
\ee where $\mt_k$ is defined in (\ref{ttt1}). Furthermore, after
some algebraic manipulations, $v_k$ is written in detail as \be
\label{vvv}
v_k=\sqrt{\frac{p_b\beta_k\bar{\vartheta}_k}{K}}(\frac{\mt^\texttt{H}_k\bar{\mH}_k^\texttt{T}\mr_k}{\sqrt{NM}})s_k
+\sqrt{\frac{p_b\beta_k\tilde{\vartheta}_k}{K}}(\frac{\mt^\texttt{H}_k\tilde{\mH}_k^\texttt{T}\mr_k}{\sqrt{NM}})s_k
+\sqrt{\frac{p_b\beta_k}{K}}\sum_{i=1,i\neq k}^K(\frac{\mt^\texttt{H}_k\mG_k^\texttt{T}\mr_i}{\sqrt{NM}})s_i+
(\frac{\mt^\texttt{H}_k\mz_k}{\sqrt{N}}). \ee

\begin{Proposition} \label{P5}
For the $k$-th user, its output statistical SINR is given by \be
\bar{\gamma}_k(K)=\frac{\frac{p_bMN}{K}\beta_k\bar{\vartheta}_k}{1+\frac{p_bN}{KM}\beta_k\bar{\vartheta}_k\sum_{i=1,i\neq
k}^K|\varrho_{ki}|^2+p_b\tilde{\vartheta}_k\beta_k} \ee
\end{Proposition}

{\em Proof:} From (\ref{vvv}) it follows that \be
\bar{\gamma}_k(K)=\frac{\frac{p_b\beta_k\bar{\vartheta}_k}{K}|\frac{\mt^\texttt{H}_k\bar{\mH}_k^\texttt{T}\mr_k}{\sqrt{NM}}|^2}
{\frac{p_b\beta_k\bar{\vartheta}_k}{K}\sum_{i=1,i\neq
k}^K|\frac{\mt^\texttt{H}_k\bar{\mH}_k^\texttt{T}\mr_i}{\sqrt{NM}}|^2+\frac{p_b\beta_k\tilde{\vartheta}_k}{K}\sum_{i=1}^K\mathbb{E}
|\frac{\mt^\texttt{H}_k\tilde{\mH}_k^\texttt{T}\mr_i}{\sqrt{NM}}|^2
+\mathbb{E}|\frac{\mt^\texttt{H}_k\mz_k}{\sqrt{N}}|^2}. \ee Thus
the desired result can be obtained under the help of the following expressions \be
\frac{1}{\sqrt{NM}}\mt^\texttt{H}_k\bar{\mH}_k^\texttt{T}\mr_k=\sqrt{NM},
\ee \be \frac{1}{\sqrt{N}}\mt^\texttt{H}_k\mz_k \sim
\texttt{CN}(0,1), \ee \be \label{ttt2}
\frac{1}{\sqrt{NM}}\mt^\texttt{H}_k\tilde{\mH}_k^\texttt{T}\mr_i
\sim \texttt{CN}(0,1), \ee and \be
\frac{1}{\sqrt{NM}}\mt^\texttt{H}_k\bar{\mH}_k^\texttt{T}\mr_i=\sqrt{\frac{N}{M}}\varrho_{ki}
\ee where $
\varrho_{ki}=\frac{1-e^{jM\varphi_{ki}}}{1-e^{j\varphi_{ki}}},$
$\varphi_{ki}=2\pi d (\sin(\theta_i)-\sin(\theta_k))$, and
$\theta_i\neq \theta_k. $ \;\; $\blacksquare$

For the asymptotic SINR analysis, we have the following results.

\begin{Corollary} \label{C7} When $M \to \infty$, let $E_b=MNp_b/K$ be fixed and $K$ be finite.
Then \be \lim_{M \to \infty}\bar{\gamma}_k(K)=\lim_{M \to
\infty}\bar{\gamma}_k(1)=E_b\beta_k\bar{\vartheta}_k. \ee
\end{Corollary}

\begin{Corollary} \label{C8} When $N \to \infty$, let $E_b=MNp_b/K$ be fixed, $M$ and $K$ be finite, and $K>1$.
Then \be \lim_{N \to
\infty}\bar{\gamma}_k(K)=\frac{E_b\beta_k\bar{\vartheta}_k}{1+E_bM^{-2}c_k(K)}<\lim_{N
\to \infty}\bar{\gamma}_k(1)=E_b\beta_k\bar{\vartheta}_k. \ee
where $c_k(K)=\beta_k\bar{\vartheta}_k\sum_{i=1,i\neq
k}^K|\varrho_{ki}|^2$.
\end{Corollary}

\begin{Corollary} \label{C9} Let
$E_b=MNp_b/K$ be fixed. When $M \to \infty$ and $K \to \infty$,
let $ \frac{K}{M^\alpha} \to \iota$, $\iota>0$. If $0<\alpha<1$,
Then \be \lim_{M \to \infty}\bar{\gamma}_k(K)=\lim_{M \to
\infty}\bar{\gamma}_k(1)=E_b\beta_k\bar{\vartheta}_k. \ee And if
$N$ be finite and $\alpha=1$, then \be \lim_{M \to
\infty}\bar{\gamma}_k(K)=\frac{E_b\beta_k\bar{\vartheta}_k}{1+E_b\iota
\tilde{\vartheta}_k\beta_k/N}<\lim_{M \to
\infty}\bar{\gamma}_k(1)=E_b\beta_k\bar{\vartheta}_k. \ee
\end{Corollary}

Single-cell and multi-cell massive-MIMO systems have been
investigated in time-division duplexing (TDD) mode
\cite{Marzetta}, \cite{Ngo}, \cite{Hoydis}. For the downlink
scenario under TDD mode, the channel matrix can be estimation
through uplink pilot training. Similar to the discuss given in the
uplink scenario, it can conclude that for each user its ergodic achievable
rate of the BF scheme based on FF and MMSE channel estimation is not larger than that of the BF scheme only based on the
specular component when the number of the BS antennas is large
enough, namely,
\begin{Theorem} \label{P6} Let $E_b=MNp_b/K$
be fixed when $M \to \infty$, and let $p_p=p_b/K$.  Then \be  \lim_{M \to
\infty}\hat{R}_k (K)=\frac{T-\tau}{T} \lim_{M \to \infty}
\bar{R}_k (K)=\frac{T-\tau}{T}\bar{R}^{\texttt{L}}_k.\ee
\end{Theorem}

\subsection{Effect of Spacial Antenna Correlation}
The aforementioned study is only limited in the spatially
uncorrelated Rician fading model, which implies a rich scattering
assumption. However, the rich scattering assumption is not always
realistic and spatial correlation comes often into play. It has
been shown that spatial antenna correlation changes drastically
with the scattering environment, the distance between the
transmitter and the receiver, the antenna configurations and the
Doppler spread  \cite{Taricco}. This motivates us to consider a
spatially correlated Rician fading model and further observe if
spatial correlation makes a great impact on the system
performance.

In the single-cell uplink or downlink scenario, the subchannel
matrix $\mH_k$ for the $k$-th user is now assumed to be of
separately correlated Rician fading type, and then (\ref{hhh1})
can be rewritten as \be \label{hhh2}
\mH_k=\sqrt{\bar{\vartheta}_k}\overline{\mH}_k+\sqrt{\tilde{\vartheta}_k}\mathbf{\Sigma}^{1/2}\widetilde{\mH}_k\mathbf{\Sigma}^{1/2}_k
\ee where $\mathbf{\Sigma}$ and $\mathbf{\Sigma}_k$ are the
correlation matrices of the BS terminal and the $k$-th user
terminal, respectively, satisfying
$\texttt{Tr}(\mathbf{\Sigma})=M$ and
$\texttt{Tr}(\mathbf{\Sigma}_k)=N$.

The scattered component makes an impact on the system performance
through statistical properties of those random variables formed by
weighting the scattered component, which have been given in
(\ref{ttt2}) under the case without spacial correlation.  When
there exists spacial correlation, (\ref{ttt2}) need to be
rewritten as \be
\frac{1}{\sqrt{NM}}\mt^\texttt{H}_k{(\mathbf{\Sigma}^{1/2}\widetilde{\mH}_k\mathbf{\Sigma}_k^{1/2})}^\texttt{T}\mr_i
\sim \texttt{CN}(0,\delta^2_{ki}). \ee where \be
\delta^2_{ki}=\frac{1}{MN}\|\mr^\texttt{H}_i\mathbf{\Sigma}^{1/2}\|^2\cdot\|\mt^\texttt{H}_k\mathbf{\Sigma}^{1/2}_k\|^2.
\ee

In Section V, by numerical results we will make a comparison
between the two cases with and without spacial correlation  and
examine the effect of spacial correlation on the individual rate.

\subsection{Generalized Single-Link Model}
So far we have investigated the transmission scheme based on
transmit and receive conjugate BFs for single-cell massive MIMO
systems through rank-1 Rician flat-fading channels.  we can
generalize the system model to the scenario including many factors
such as distributed MIMO (\cite{Matthaiou},
\cite{Zhang1},\cite{Zhang2}), multi-cell (\cite{Ngo1},
\cite{Ngo2}), heterogeneous network (\cite{Jeong}), spacial
correlation (\cite{Wen}, \cite{Taricco}, \cite{Riegler},
\cite{Siriteanu}), high-rank Rician (\cite{Matthaiou1},
\cite{Bohagen}), and frequency-Selective fading (\cite{Jin},
\cite{Mckay}).  In the general scenario, a unified single-link
transceiver model can be naturally formed by taking account of
these factors, and noting such two facts that

a) A frequency selective MIMO channel can be converted into a set
of parallel independent flat MIMO channels \cite{Jin},
\cite{Mckay};

b) For two independent complex random matrices following
noncentral Gaussian distributions, their sum also follows a
noncentral Gaussian distribution.

We consider a massive-MIMO link with $N$ transmit antennas and $M$
receive antennas, where $NM$ is assumed to be very large. The $M
\times 1$ received signal vector is represented as \be
\my=\sqrt{E}\mG\mx+\sum_{i=1}^L\sqrt{E_i}\mG_i\mx_i+\mz \ee where
$\mz$  still denotes the AWGN; $\mG=\bar{\mG}+\tilde{\mG}$ denotes
the $M \times N$ desired channel matrix following Rician fading
distribution; $\mG_i$ denotes the $i$-th $M \times N_i$
interfering channel matrix with Rician fading distribution; $E$
and $E_i$ are the average transmitted powers of the desired user
and the $i$-th interference user, respectively; and  $\mx$ and
$\mx_i$ are the transmitted symbol vector for the desired user and
the $i$-th interference user, respectively.

Suppose that the transmitter and receiver only know the LOS
component $\bar{\mG}$. Then the optimal transmit/receive conjugate
BF vectors $\mb$ and $\mw$ can be given by \cite{Yue},\cite{Kang}
\be \mw=\bar{\mG}\mb, \;\; \mb=\mmu_{\texttt{max}} \ee where
$\mmu_{\texttt{max}}$ denotes eigenvector corresponding to the
largest eigenvalue $\lambda({MN})$ of the quadratic form
$\bar{\mG}^{\texttt{H}}\bar{\mG}$. We assume that $\mx_i=\mb_is_i$
with $|\mb_i^\texttt{H}\mb_i|=1$ and $\mathbb{E}\|\ms_i\|^2=1$.

This conjugate BF transmission scheme in the interference network
would be attractive in practice if the following favorable
propagation condition could be met:\be \label{eee1}
\frac{\mathbb{E}|\mw^\texttt{H}\sqrt{E}\tilde{\mG}\mb|^2}{\lambda({MN})}\leq
\varepsilon_1, \ee \be \label{eee2}
\frac{\sum_{i=1}^L\mathbb{E}|\mw^\texttt{H}\sqrt{E_i}\mG_i\mb_i|^2}{\lambda({MN})}\leq
\varepsilon_2 ,\ee and \be \label{eee3}
\frac{\mathbb{E}|\mw^\texttt{H}\mz|^2}{\lambda({MN})}\leq
\varepsilon_3 \ee where $\varepsilon_1$, $\varepsilon_2$,  and
$\varepsilon_3$ are parameters involving quality of service (QoS)
given in advance. By (\ref{eee1}), (\ref{eee2}) and (\ref{eee3}),
the output statistical SINR at the receiver is bounded by \be \bar{\gamma} \geq
\frac{E}{\varepsilon_1+\varepsilon_2+\varepsilon_3}.\ee The
asymptotic analysis can be carried on by following a similar line
of reasoning as in the single-cell case.

\section{Simulation Results}
In this section, we present analytical results and simulation
results for a single-cell with a radius of $r_m=1000$ meters and
$K$ users distributed randomly and uniformly over the cell. It is
assumed that there no user is closer to the BS than $r_h=100$
meters. We first study the simple single-user case and then the
uplink and downlink cases with multi-users. In all simulation, the
large-scale fading coefficient $\beta_k$ for the $k$-th user (or
$\beta$) is always modelled as \cite{Ngo}, \cite{Zhang},\be
\beta_k=z_k/(r_k/r_h)^v \ee where $z_k$ is a log-normal random
variable with standard deviation $\sigma=8 \mbox{dB}$, $v$ denotes
the path loss exponent and is set to be $3.8$, and $r_k \in
[r_h,r_m]$ denotes the distance between the underlying user and
the BS. Note that $\mathbb{E}\beta_k=\bar{\beta}_k=0.20479.$ For
simplicity, we always assume that all users have the identical
Rician factor. Unless otherwise stated, the antenna spacings are
assumed to be $d=d_r=0.3$ and $d_k=d_t=0.3$.

The lower bound of individual ergodic achievable rate $\bar{R}_k(K)$
is quite tight, especially at large $M$. Therefore, in the following,
we will use $\bar{R}_k(K)$ to replace the exact rate $R_k(K)$ for all numerical work.

\subsection{Single-user Massive-MIMO systems}
\begin{figure}[t]
\centering
\includegraphics[width=120mm]{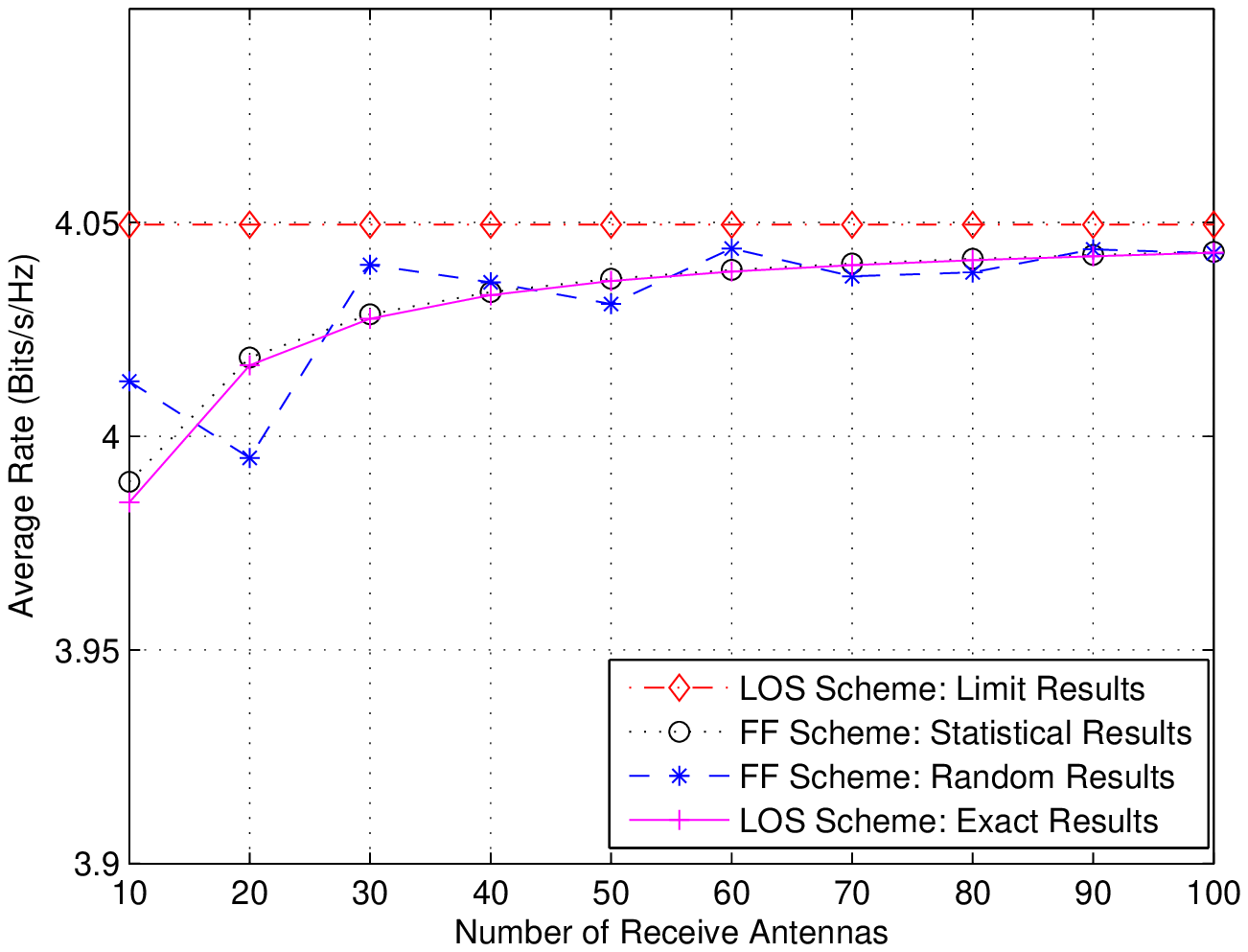}
\caption{Average rate versus the number of receiver antennas in
the single user scenario for making a comparison between the LOS-
and FF-based schemes when the large-scale fading parameter is
fixed.} \label{Fig1}
\end{figure}

\begin{figure}[t]
\centering
\includegraphics[width=120mm]{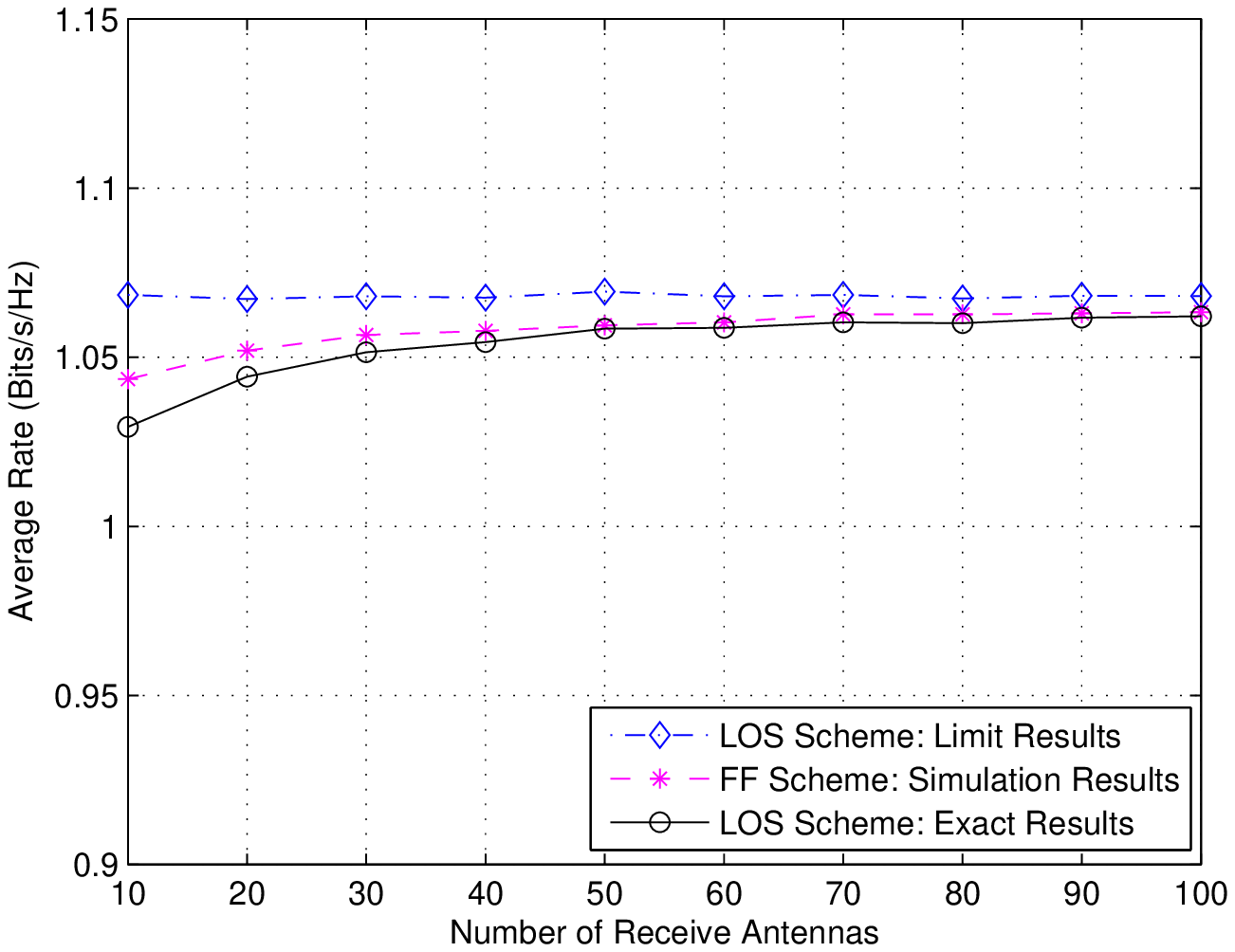}
\caption{ Average rate versus the number of receiver antennas in
the single user scenario for making a comparison between the LOS-
and FF-based schemes when the large-scale fading parameter is
random.} \label{Fig2}
\end{figure}

We first validate Theorem \ref{P2} by Figure \ref{Fig1} and
\ref{Fig2}. If the transmitted data are modulated with orthogonal
frequency division multiplexing (OFDM), as in\cite{Ngo} and
\cite{Zhang}, the coherence time of the channel can be chose to be
$T=196$ according to the LTE standard, and the length of uplink
pilots can be chosen as $\tau=K$ for MMSE channel estimation. So
$\frac{T-\tau}{T}=195/196$. In order to make a convenient
comparison, however, below we consider to compare $\bar{R}$ with
$\hat{R}^{\tau}=\frac{T}{T-\tau}\hat{R}$ rather than $\hat{R}$
when the large-scale fading coefficient and the angle of arrival
are set to be $\beta=\bar{\beta}=0.20479$ and $\theta=\pi/4$,
respectively.

Assuming that $N$=10, $E_u=20 \mbox{dB}$ and $\vartheta=5
\mbox{dB}$, Figure \ref{Fig1} plots $\bar{R}$ and
$\hat{R}^{\tau}$ as two functions of the number of receive
antennas $M$. For further comparison, Figure \ref{Fig1} also plots
the rate limit $\bar{R}^\texttt{L}$ and a random example of the
instantaneous rate when the FF-based scheme is used. As expected,
Figure \ref{Fig1} shows such a power scaling law that as $M$
grows large, both of the rates $\bar{R}$ and $\hat{R}^{\tau}$ tend to
the rate limit $\bar{R}^\texttt{L}$.  It can be seen from this
figure that the two rates are close to each other when $M$ is
relatively large while $\hat{R}^{\tau}$ is higher than $\bar{R}$
when $M$ is relatively small.  Moreover, the random rate curve
has a large fluctuation, but still tends to the limit result when
$M$ is very large.

Furthermore, in the case where both $\beta$ and $\theta$ are
assume to be random, Figure \ref{Fig2} plots $\mathbb{E}\bar{R}$
and $\mathbb{E}\hat{R}^{\tau}$ as the number of receive antennas
$M$ changes. The behavior of the two rates  with LOS- and FF-based
schemes is similar to that in the case where $\beta$ and $\theta$
are fixed. From Figure \ref{Fig1} and Figure \ref{Fig2}, however,
it can be found that the following approximation result does not
hold \be \mathbb{E}\bar{R} \approx \log_2(1+\mathbb{E}(E_u
\beta\bar{\vartheta}))=\log_2(1+(E_u
\bar{\beta}\bar{\vartheta})).\ee

\begin{figure}[t]
\centering
\includegraphics[width=120mm]{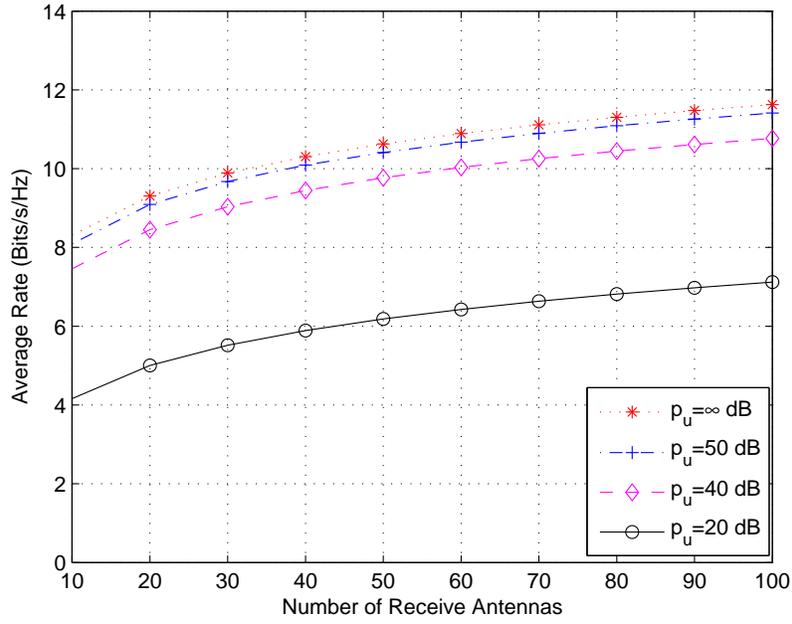}
\caption{ Average rate versus the number of receiver antennas in
the single user scenario for different transmit powers.}
\label{Fig3}
\end{figure}

\begin{figure}[t]
\centering
\includegraphics[width=120mm]{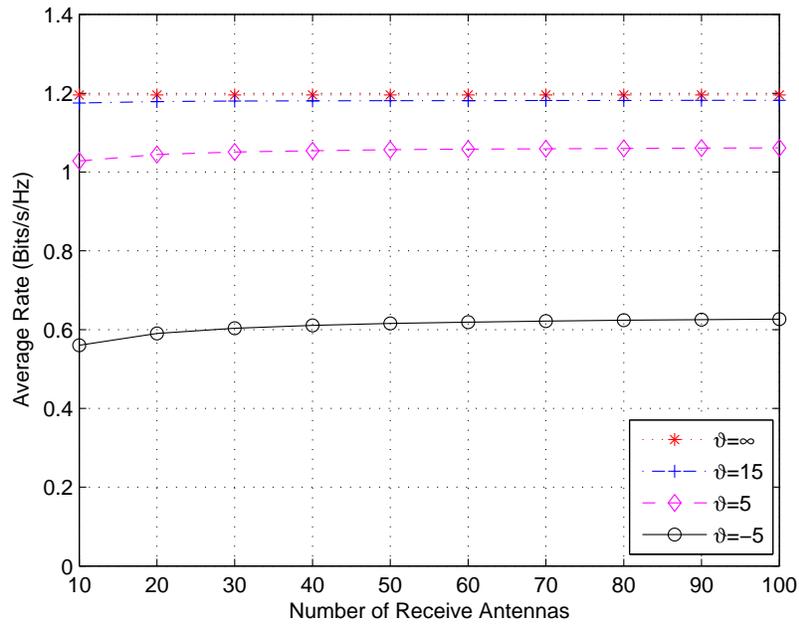}
\caption{ Average rate versus the number of receiver antennas in
the single user scenario for different Rician factors.}
\label{Fig4}
\end{figure}

\begin{figure}[t]
\centering
\includegraphics[width=120mm]{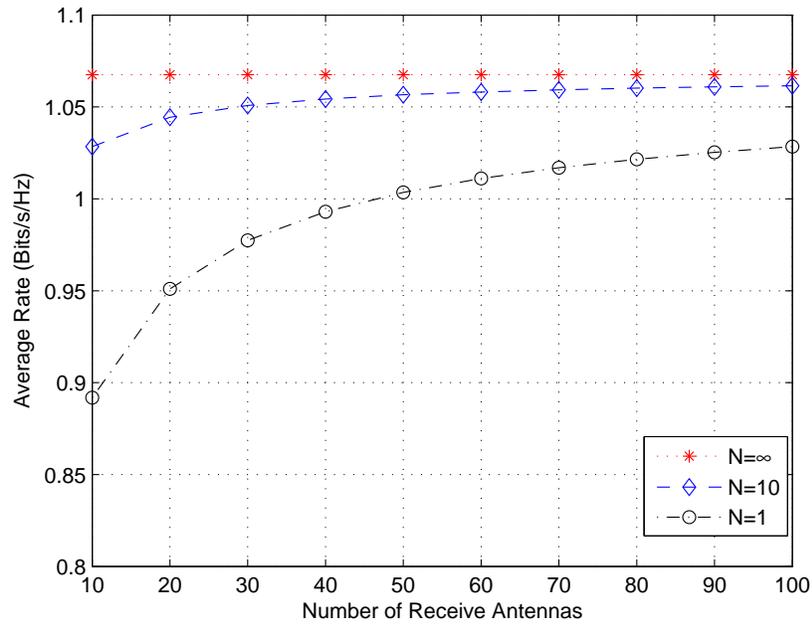}
\caption{ Average rate versus the number of receiver antennas in
the single user scenario for different numbers of transmit
antennas.} \label{Fig5}
\end{figure}

When $\beta$ is random, we next consider to verify Proposition 1
and its corollaries by Figure \ref{Fig3}, Figure \ref{Fig4} and
Figure \ref{Fig5}. Under the case of $N=10$ and $\vartheta=5
\mbox{dB}$, Figure \ref{Fig3} shows that although
$\mathbb{E}\bar{R}$ improves with an increasing $p_u$, the amount
of improvement becomes smaller and smaller. In Figure \ref{Fig4},
we set $N=10$ and $E_u=20 \mbox{dB}$. It can be observed from
Figure \ref{Fig4} that the average rate $\mathbb{E}\bar{R}$
increases as $\vartheta$ increases,  but the amount of rate
improvement becomes smaller and smaller. In Figure \ref{Fig5}, we
assume that $\vartheta=5 \mbox{dB}$ and $E_u=20 \mbox{dB}$.
Although the two rates with $N=10$ and $N=1$ can increase as $M$
grows from $10$ to $100$, only the one with $N=10$ is close to the
theoretical limit when $M=100$. Moreover, it is interestingly
found in Figure \ref{Fig5} that the rate result with $N=10$ and
$M=10$ is equal to the rate result with $N=1$ and $M=100$. This
phenomena can be explained by Corollary 3, and gives us a
suggestion for practical MIMO configuration.

\subsection{Uplink Massive-MIMO systems}
\begin{figure}[t]
\centering
\includegraphics[width=120mm]{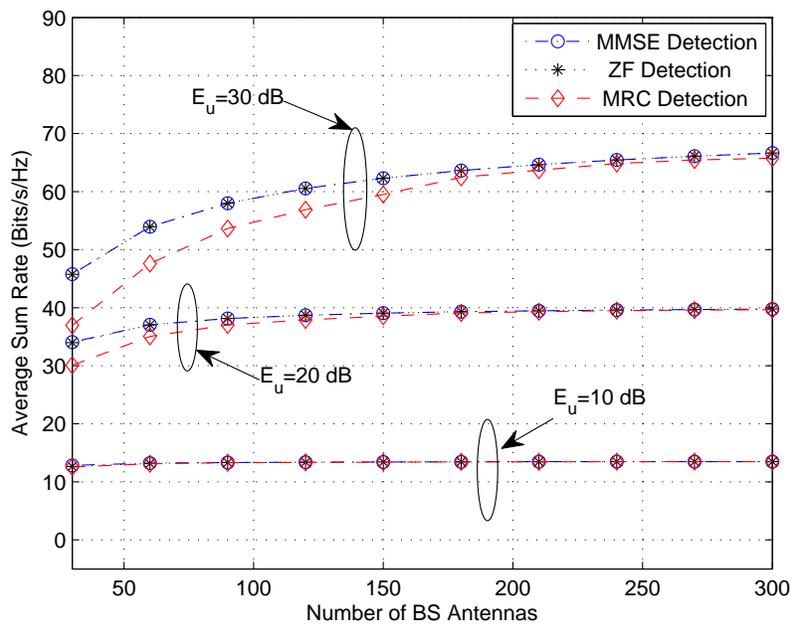}
\caption{Average sum rate versus the number of BS antennas in the
uplink scenario for making a comparison among BF, ZF and MMSE
linear detectors when several values of $E_u$ are used.}
\label{Fig6}
\end{figure}

We now turn our attention to the uplink scenario, and first make a
comparison among the MMSE, ZF, and MRC linear detection schemes.
When $N=3$, $K=10$, $\vartheta_k=5 \mbox{dB}$, $\beta_k=0.20479$,
and $k=1,2,\ldots, K$, Figure \ref{Fig6} plots three curves about
the average sum rate for $E_u=30, 20, 10 \mbox{dB}.$  As expected,
simulation results show that the MMSE scheme is always optimal
among them. When $M \geq 30$, however, the ZF scheme and the MMSE
scheme can perform similarly in the terms of the sum rate, and
outperform the MRC scheme for $E_u=30\mbox{dB}$ and
$E_u=20\mbox{dB}$. When $E_u=10\mbox{dB}$, all of these three
scheme have similar performance. This means that the MRC tends to
the optimal MMSE with a decreasing $E_u$.

\begin{figure}[t]
\centering
\includegraphics[width=120mm]{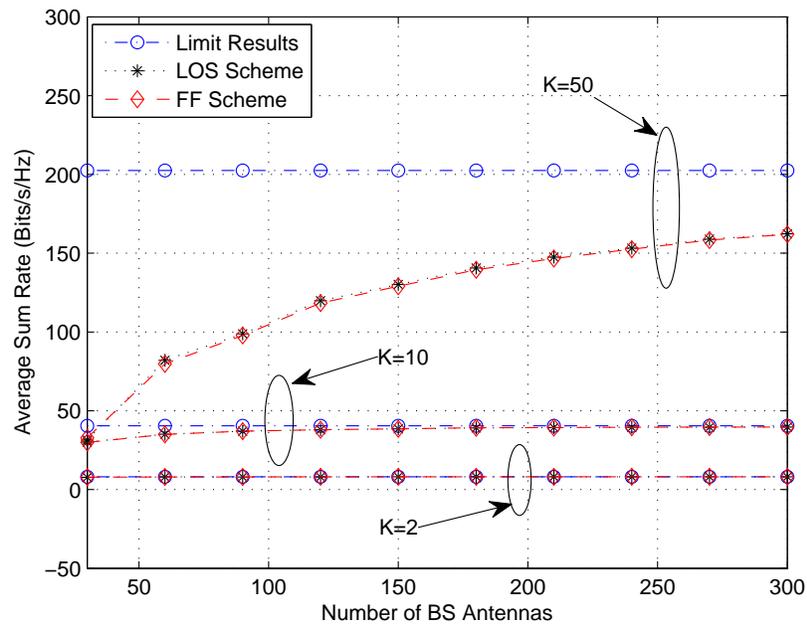}
\caption{ Average sum rate versus the number of BS antennas in the
uplink scenario for making a comparison between the LOS- and
FF-based schemes when several values of $K$ are used.}
\label{Fig7}
\end{figure}

We set $N=3$, $E_u=20 \mbox{dB}$, $\vartheta_k=5 \mbox{dB}$,
$\beta_k=0.20479$, and $k=1,2,\ldots, K$. In this case, we
consider to corroborate Theorem \ref{P4} by comparing the sum
rate of the LOS-based scheme with that of the FF-based scheme. For
$K=50, 10, 2$, Figure \ref{Fig7} plots the two sum rates of
$\sum_{k=1}^K\bar{R}_k(K)$ and
$\sum_{k=1}^K\hat{R}_k^{\tau}(K)=\frac{T}{T-\tau}\sum_{k=1}^K{\hat{R}}_k(K)$
as $M$ increases from $30$ to $300$ . For comparison, three curves
are also plotted by using the corresponding limit results
$K\bar{R}^\texttt{L}$. From Figure \ref{Fig7} it can be observed
that the two LOS- and FF-based schemes perform similarly, and
their sum rates tend to the corresponding limit results for the
three different values of $K$. However, as $K$ increases, the
inter-user interference becomes stronger and stronger, and thus
the speed of convergence becomes slower and slower even although
$M$ is very large.

\begin{figure}[t]
\centering
\includegraphics[width=120mm]{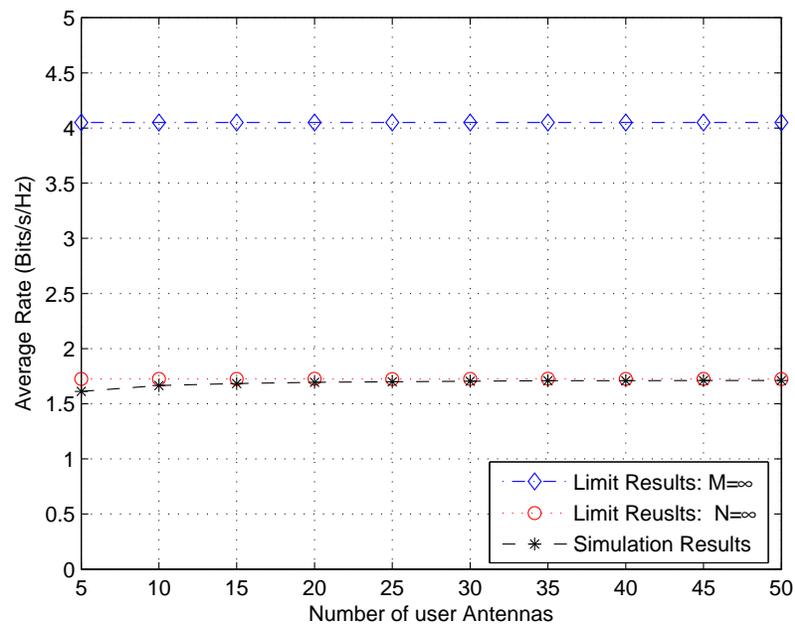}
\caption{ Average individual rate versus the number of user
antennas in the uplink scenario.} \label{Fig8}
\end{figure}

It should be noticed that in Figures \ref{Fig6} and \ref{Fig7} the
angle of arrival for the $k$-th user is set to be
$\theta_k=\frac{\pi}{2}+\frac{2k-1}{2K}, $ $k=1,2,\ldots, K$.

we examine Corollary \ref{C5} by Figure \ref{Fig8}, assuming that
the angle of arrival for the $k$-th user $\theta_k$ is random. We
further assume that $M=20$, $K=10$, $E_u=20 \mbox{dB}$,
$\vartheta_k=5 \mbox{dB}$, $\beta_k=0.20479$, and $k=1,2,\ldots,
K$. Different from the single-user case, the individual rate limit
with $M \to \infty$ is obviously higher than the one with $N \to
\infty$. Even when $N$ is not very large, the exact individual
rate can be quite close to the latter.

\subsection{Impacts of the number of users, correlation coefficients, and antenna spacings}
\begin{figure}[t]
\centering
\includegraphics[width=120mm]{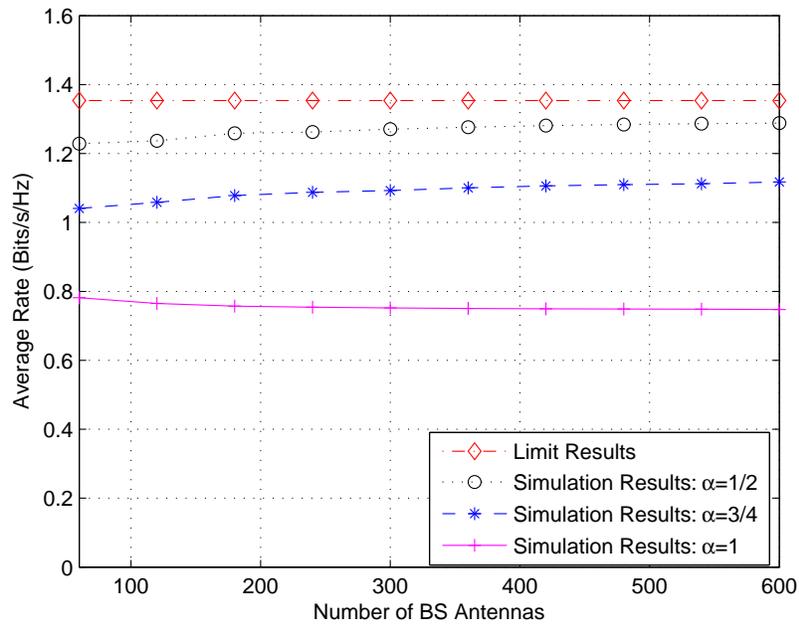}
\caption{Average individual rate versus the number of BS antennas
in the downlink scenario for different values of the parameter
$\alpha$.} \label{Fig9}
\end{figure}

\begin{figure}[t]
\centering
\includegraphics[width=120mm]{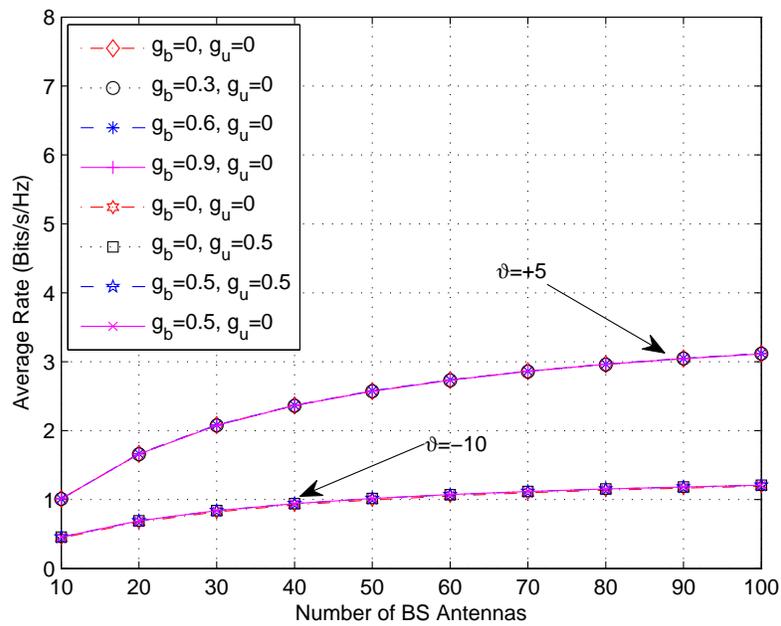}
\caption{Average individual rate versus the number of BS antennas
in the downlink scenario for various pairs of spacial correlation
coefficients.} \label{Fig10}
\end{figure}

\begin{figure}[t]
\centering
\includegraphics[width=120mm]{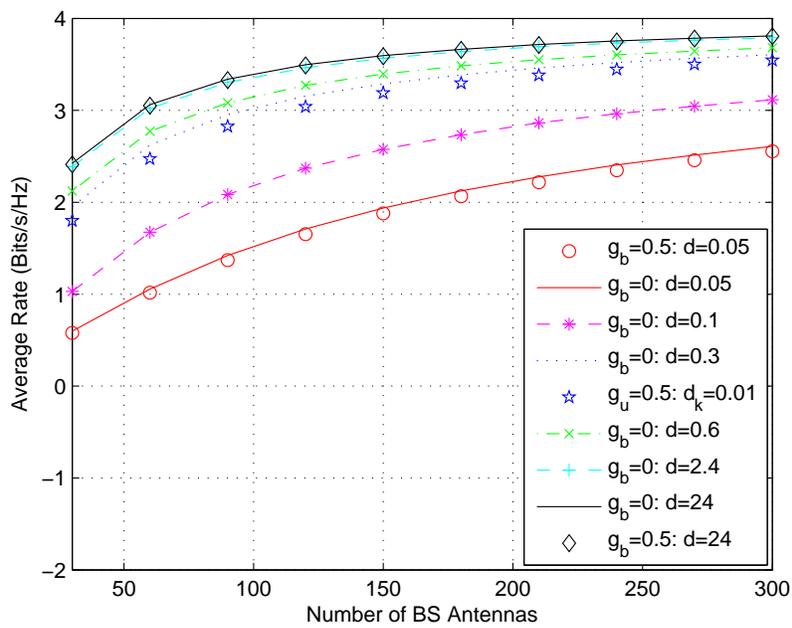}
\caption{Average individual rate versus the number of BS antennas
in the downlink scenario for various values of antenna spacing.}
\label{Fig11}
\end{figure}

Finally, we conduct experiments for the downlink scenario and
analyze the impact of the number of users, correlation
coefficients, and antenna spacings on the rate performance. In
Figures \ref{Fig9}-\ref{Fig11}, we assume that the angle of
arrival for the $k$-th user $\theta_k$ is random.

In order to verify Corollary \ref{C9} in Section IV, Figure
\ref{Fig9} plots the average individual rate as a function of the
number of BS for three different values of $\alpha=1/2, 3/4, 1$
under the case with $\iota=1/2$. In Figure \ref{Fig9}, we assume
that $N=2$, $E_u=10 \mbox{dB}$, $\vartheta_k=5 \mbox{dB}$,
$\beta_k=0.20479$, and the rate limit is included for comparison.
As expected, when $\alpha$ decreases, the number of users also
decreases, and the average individual rate be closer to the rate
limit.  On the other hand, for a fixed $\alpha \in (0, 1)$ the
average individual rate can improve as the number of BS antennas
grows large from $60$ to $600$, but the amount of improvement is
extremely small.  When $\alpha$ is fixed to be one, however, the
average individual rate keeps almost unchange as the number of BS
antennas grows large from $200$ to $600$.

In what follows, we observe by simulation the impact of spacial
correlation and antenna array structure on the achievable rate in
the downlink scenario.

In order to provide an assessment of the influence of spacial
correlation, Figure \ref{Fig10} plots the average individual rate
as a function of the number of BS antennas for various spacial
correlation cases when $\beta_k=0.20479$, $K=10$, $E_b=20
\mbox{dB}$ and $N=10$. In Figure \ref{Fig10}, we employ
exponential correlation model such that the BS correlation matrix
and the user correlation matrix can be expressed as
$(g_b^{|i-j|})$ and $(g_u^{|i-j|})$ where $g_b \in [0,1]$ and $g_u
\in [0,1]$ denote their correlation coefficients. Note that
$g_b=0$ (or $g_u=0$) indicate that there is no spacial correlation
at the BS antenna array (or user antenna array). It can be clearly
seen from Figure \ref{Fig10} that when $\vartheta_k=5 \mbox{dB}$,
the four rate curves plotted for four different values of the BS
correlation coefficient $g_b=0, 0.3, 0.6, 0.9$ are quite similar
under the case of $g_u=0$.  This means that the impact of the BS
spacial correlation on the rate can be negligible. For
$\vartheta_k=-10 \mbox{dB}$, Figure \ref{Fig10} also plots four
rate curves which correspond the four cases as follows: i) the
uncorrelated case of $(g_b=0, g_u=0)$; ii) the semi-correlated
case of $(g_b=0, g_u=0.5)$; iii) the semi-correlated case of
$(g_b=0.5, g_u=0)$, and iv) the double-correlated case of
$(g_b=0.5, g_u=0.5)$. The four rate curves are obviously similar.
This implies that neither the BS spacial correlation nor the user
spacial correlation make a serious impact on the rate performance.

To examine the influence of antenna spacings on the rate
performance of uncorrelated MIMO systems, Figure \ref{Fig11}
depicts the average individual rate against the number of BS
antennas for various antenna spacing cases when $\beta_k=0.20479$,
$d_k=0.3$, $K=10$, $N=3$, and $E_b=20 \mbox{dB}$. It can be
clearly seen from Figure \ref{Fig11} that as $d$ grows large from $0.05$
to $24$, the average individual rate increases, but the amount of
increasing gradually becomes smaller. In particular, the rate values
with $d=2.4$ is quite close to the ones with $d=24$. It is
interesting to notice that the individual rate with $d=0.05$ and
$M=300$ is an accurate approximation of that with $d=2.4$ and
$M=30$. This example implies such a power scaling law that the
transmit power needed in the former case is one tenth of the one
needed in the latter case.

For comparison with the correlation case, Figure \ref{Fig11} also
depicts the average individual rate against $M$ for the following
three correlation cases: i) $g_b=0.5$ and $g_u=0$ when $d=0.05$
and $d_k=0.3$; ii) $g_b=0.5$ and $g_u=0.5$ when $d=0.3$ and
$d_k=0.01$; iii) $g_b=0.5$ and $g_u=0$ when $d=24$ and $d_k=0.3$.
Similar to the case with $d=0.3$ and $d_k=0.3$ shown in Figure
\ref{Fig10}, it concludes that spacial correlation has no serious
influence on the rate performance.

\section{Conclusion}
In this paper, we have investigated the LOS-based conjugate BF
transmission scheme for massive-MIMO systems over Rician fading
channels, derived some expressions of the statistical SINR, and presented several
power scaling laws for several cases under the help of these expressions. By
comparison with pure Rayleigh fading environments, it can be found
that massive MIMO systems are more suitable to be deployed in
Rician fading environments, especially when the LOS component is
strong.

Numerical results have been conducted only by the system models
with some ideal assumptions. In future work, we will examine our
theoretical analysis based on more practical system models.


\end{document}